\documentclass[aps,proof]{revtex4-1}%
\usepackage{amssymb}
\usepackage{amsfonts}
\usepackage{amsmath}
\usepackage{graphicx}%
\begin{document}

\title{Direct correlation function from the consistent fundamental-measure free energies for hard-sphere mixtures''}
\author{James F. Lutsko}
\affiliation{Center for Nonlinear Phenomena and Complex Systems, Code Postal 231,
Universit\'{e} Libre de Bruxelles, Blvd. du Triomphe, 1050 Brussels, Belgium}
\email{jlutsko@ulb.ac.be}
\homepage{http://www.lutsko.com}

\begin{abstract}
In a recent publication[PRE 86, 04012 (2012)], Santos has presented a self-consistency condition that can be used to limit the possible forms of Fundamental Measure Theory. Here, the direct correlation function resulting from the Santos functional is derived and it is found to diverge for all densities.
\end{abstract}

\date{\today }

\maketitle

In a recent contribution, Santos introduced a novel argument aimed at
eliminating a source of ambiguity in the derivation of the Fundamental Measure
Theory (FMT) approach to Density Functional Theory for hard spheres\cite{Santos}. The
result is a new ansatz for improvement of FMT beyond the basic Rosenfeld
functional\cite{rosenfeld1}. The proposal is quite interesting as the most accurate density
functionals currently in use (such as the "White Bear" functional\cite{white_bear,white_bear_2}) are of
exactly this type:\ heuristic improvements beyond functionals based on
Rosenfled's original reasoning together with the additional requirement that
the forms reproduce known, exact results in low-dimensional systems. The
introduction of a new element that eliminates some of the arbitrarity of these
extensions is therefore welcome. The proposal of Santos is based on an exact
scaling relation of the type successfully exploited by him and co-workers in
the development of highly accurate approximations for the free energy and pair-distribution
function of mixtures of hard spheres\cite{Santos1, Santos2}. The purpose of this Comment is to
examine one consequence of the proposed ansatz, namely the implied direct
correlation function (DCF).

The direct correlation function  is a fundamental element in DFT as it provides a connection between model free energy functionals and liquid-state properties, for which much is known\cite{LutskoAdvChemPhysDFT}. Given a
(grand-canonical) free energy functional, $\Omega\left[  \rho\right]
=F_{id}\left[  \rho\right]  +F_{ex}\left[  \rho\right]  -\mu\rho$, where
$\rho\left(  \mathbf{r}\right)  $, is the ensemble-averaged local density,
$F_{id}$ is the ideal gas contribution, which is not relevant here, $\mu$ is
the chemical potential and $F_{ex}\left[  \rho\right]  $ is the excess term,
the (two-body) direct correlation function is given by taking two functional
derivatives with respect to the density,
\begin{equation}
c_{2}\left(  \mathbf{r}_{1},\mathbf{r}_{2}\right)  =-\frac{\delta^{2}\beta
F_{ex}\left[  \rho\right]  }{\delta\rho\left(  \mathbf{r}_{1}\right)
\delta\rho\left(  \mathbf{r}_{2}\right)  },\label{1}%
\end{equation}
where $\beta=1/k_{B}T$ , $k_{B}$ is Boltzmann's constant and $T$ is the
temperature\cite{Evans79,LutskoAdvChemPhysDFT}. This relation between the free energy functional and the
DCF has always provided an important connection between free energy models and
liquid-state properties:  for example, one of the first indications of the
utility of the White Bear functional was its improvement in the predicted DCF
of hard spheres\cite{white_bear}. 

In DFT, the only unknown is the excess term and FMT\ is based on an ansatz of
the form%
\begin{equation}
\beta F_{ex}=\int\Phi\left(  \mathbf{n}\left(  \mathbf{r;}[\rho]\right)
\right)  d\mathbf{r}\label{2}%
\end{equation}
where the weighted densities have the generic expressions%
\begin{equation}
n_{i}\left(  \mathbf{r;[\rho]}\right)  =\int w_{i}\left(  \mathbf{r-r}%
^{\prime}\right)  \rho\left(  \mathbf{r}^{\prime}\right)  d\mathbf{r}^{\prime}%
\end{equation}
Different models involve different collections of density-independent weight
functions, $w_{i}$, and of different forms for the function $\Phi\left(
\mathbf{n}\right)  $. The proposal of Santos makes use of the weight functions
as were introduced by Rosenfeld ($w_{s}\left(  \mathbf{r}_{12}\right)
=\delta\left(  \frac{\sigma}{2}-r_{12}\right)  $, $w_{\eta}\left(
\mathbf{r}_{12}\right)  =\Theta\left(  \frac{\sigma}{2}-r_{12}\right)  $,
$w_{v_{i}}\left(  \mathbf{r}_{12}\right)  =\widehat{r}_{12,i}\delta\left(
\frac{\sigma}{2}-r_{12}\right)  $ where $\sigma$ is the hard-sphere diameter)
and the  $\Phi$ function  of the Rosenfeld, $\Phi_{R}=s\Phi_{1}\left(
\eta\right)  +\Phi_{2}\left(  \eta\right)  \left(  s^{2}-v^{2}\right)
+\Phi_{3}\left(  \eta\right)  s\left(  s^{2}-3v^{2}\right)  $  , where
$\eta\left(  \mathbf{r};[\rho]\right)  =n_{\eta}\left(  \mathbf{r;[\rho
]}\right)  $ is the weighted density formed from the weight function $w_{\eta
}\left(  \mathbf{r}_{12}\right)  $, etc. The other terms are
\begin{eqnarray}
&\Phi_{1}\left(  \eta\right)  =-\frac{1}{\pi\sigma^{2}}\ln\left(
1-\eta\right)  ,\;\Phi_{2}\left(  \eta\right)  =\frac{1}{2\pi\sigma}\frac
{1}{\left(  1-\eta\right)  },\\
&\Phi_{3}\left(  \eta\right)  =\frac{1}{24\pi
}\frac{1}{\left(  1-\eta\right)  ^{2}} \notag
\end{eqnarray}
The Rosenfeld functional reproduces the Percus-Yevik equation of state for a
uniform fluid ($\rho\left(  \mathbf{r}\right)  =\rho\mathbf{\ }$with $\rho$
being a position-independent constant)\cite{rosenfeld1, LutskoAdvChemPhysDFT} (see the Appendix for details). The idea of the extension discussed by
Santos is that one would like to use knowledge of more accurate equations of
state than Percus-Yevik to construct potentially more accurate approximations
for $\Phi$. Note that for a uniform system, Eq.(2) shows that the excess free
energy of a uniform liquid is simply $\beta F_{ex}=V\Phi\left(  \mathbf{n}%
\left(  \rho\right)  \right)  $ where $V$ is the total volume.  Santos
introduces a correction so that $\Phi=\Phi_{R}+\Phi_{S}$ the form of which is
fixed by the scaling relations to be%
\[
\Phi_{S}\left(  \mathbf{n}\right)  =\left(  s^{2}-v^{2}\right)  \Phi
_{2}\left(  \eta\right)  \Delta\left(  \frac{2s\left(  s^{2}-3v^{2}\right)
}{\left(  s^{2}-v^{2}\right)  }\frac{\Phi_{3}\left(  \eta\right)  }{\Phi
_{2}\left(  \eta\right)  }\right)
\]
\begin{figure}[htb]
\includegraphics[angle=-90,scale=0.4]{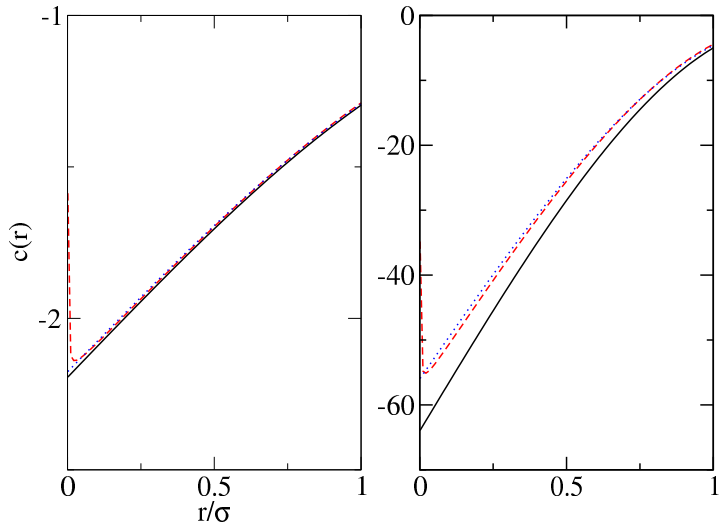} 
\caption{The direct correlation function as calculated using the Percus-Yevik approximation, full line, the White-Bear approximation, dotted line, and Eq.(\ref{result}) with $\Delta(y) =(1-(y/2)-log(1+y)/y)/3$\cite{Santos}. The left panel is for a low-density fluid with packing fraction $\eta = 0.1$ and the right panel is for a high-density fluid with $\eta = 0.5$. At low densities, the Percus-Yevik result is quite accurate and at high densities, the White-Bear result is in good agreement with simulations\cite{white_bear}.} \label{fig1}
\end{figure}

Here, the function $\Delta\left(  y\right)  $ is chosen so that the free
energy in the uniform limit agrees with some chosen form (such as
Carnahan-Starling\cite{Carnahan-Starling}). Calculation of the implied DCF in the uniform state based on Eq.(1) is
straightforward (details are given in the Appendix). The result
is
\begin{widetext}
\begin{align} \label{result}
c\left(  r;\rho\right)   &  =c_{PY}\left(  r;\rho\right)  -\frac{1}{1-\eta
}\left(  x+\frac{6\eta}{1-\eta}\left(  1-x\right)  +\frac{3\eta^{2}}{\left(
1-\eta\right)  ^{2}}\left(  x-1\right)  ^{2}\left(  x+2\right)  \right)
\Delta\left(  \frac{\eta}{1-\eta}\right)  \\
&  -\frac{2\eta}{\left(  1-\eta\right)  ^{2}}\left(  x+\frac{6\eta}{1-\eta
}\left(  1-x\right)  +\frac{3}{2}\eta^{2}\frac{2-\eta}{\left(  1-\eta\right)
^{2}}\left(  x-1\right)  ^{2}\left(  x+2\right)  \right)  \Delta^{\prime
}\left(  \frac{\eta}{1-\eta}\right)  \nonumber\\
&  -\frac{\eta^{2}}{4\left(  1-\eta\right)  ^{3}}\left(  \frac{1}{x}%
+\frac{12\eta}{1-\eta}\left(  1-x\right)  +\frac{6\eta^{2}}{\left(
1-\eta\right)  ^{2}}\left(  x-1\right)  ^{2}\left(  x+2\right)  \right)
\Delta^{\prime\prime}\left(  \frac{\eta}{1-\eta}\right)  \nonumber
\end{align}
\end{widetext}
where $x=r/\sigma$ and $\eta=\frac{\pi}{6}\rho\sigma^{3}$ and $c_{PY}\left(
r;\rho\right)  $ is the Percus-Yevik DCF that comes from $\Phi_{R}$. One
feature that stands out is that this function diverges for $x=0$ unless
$\Delta^{\prime\prime}\left(  \frac{\eta}{1-\eta}\right)  =0$. This divergence
is unphysical and, as shown in the Figure, spoils the otherwise reasonable agreement with the results of the White-Bear functional. Hence, if this undesirable behavior  is to be avoided, the only possibility is the relatively
restricted set of corrections given by $\Delta\left(  y\right)  =a+by$. However, Santos notes that in general one expects that $\Delta(y) \sim O(y^2)$ so this eliminates the possibility of a correction. Santos also offers a modified version of his proposal that appears to avoid the divergence in the DCF but at the cost of violating his self-consistency condition\cite{Santos}. It is interesting to note that similar terms arise in
deriving the DCF from the Rosenfeld part of the free energy, but they cancel
thus leaving the (finite) Percus-Yevik result (for a demonstration, see, e.g., the Appendix).

In summary, the proposal of Santos produces a divergent DCF in the uniform
liquid. Nevertheless, the reasoning behind it seems sound and it is to be
hoped that there might still be a way to exploit it so as to eliminate some of the
ambiguities of the usual extensions of FMT while retaining their advantages, one of which is an excellent description of the DCF for hard-spheres.

\begin{acknowledgments}
This work was partially supported in part by the European Space Agency under
contract number ESA AO-2004-070.
\end{acknowledgments}

\appendix{}

\section*{Calculation of the direct correlation function}

\subsection{The functional derivatives}

In general, 
\begin{eqnarray}
\frac{\delta \beta F_{ex}}{\delta \rho \left( \mathbf{r}_{1}\right) } &=&%
\frac{\delta }{\delta \rho \left( \mathbf{r}_{1}\right) }\int \Phi\left( 
\mathbf{n}\left( \mathbf{r}\right) \right) d\mathbf{r} \\
&=&\int \left( \frac{\partial }{\partial n_{i}}\Phi\left( \mathbf{n}\right)
\right) _{\mathbf{n}\left( \mathbf{r}\right) }\frac{\delta n_{i}\left( 
\mathbf{r}\right) }{\delta \rho \left( \mathbf{r}_{1}\right) }d\mathbf{r} 
\notag \\
&\equiv &\int \Phi_{i}\left( \mathbf{n}\left( \mathbf{r}\right) \right)
w_{i}\left( \mathbf{r-r}_{1}\right) d\mathbf{r},\;\;\Phi_{i}\equiv \frac{%
\partial }{\partial n_{i}}\Phi  \notag
\end{eqnarray}%
and%
\begin{equation}
\frac{\delta ^{2}\beta F_{ex}}{\delta \rho \left( \mathbf{r}_{1}\right)
\delta \rho \left( \mathbf{r}_{2}\right) }=\int \Phi_{ij}\left( \mathbf{n}%
\left( \mathbf{r}\right) \right) w_{i}\left( \mathbf{r-r}_{1}\right)
w_{j}\left( \mathbf{r-r}_{2}\right) d\mathbf{r},\;\;\Phi_{ij}\equiv \frac{%
\partial ^{2}}{\partial n_{i}\partial n_{j}}\Phi
\end{equation}%
so that the bulk limit is 
\begin{eqnarray}
\lim_{\rho \left( \mathbf{r}\right) \rightarrow \rho }\frac{\delta ^{2}\beta
F_{ex}}{\delta \rho \left( \mathbf{r}_{1}\right) \delta \rho \left( \mathbf{r%
}_{2}\right) } &=&\Phi_{ij}\left( \mathbf{n}\right) \int w_{i}\left( \mathbf{r-r%
}_{1}\right) w_{j}\left( \mathbf{r-r}_{2}\right) d\mathbf{r} \\
&\equiv &\Phi_{ij}\left( \mathbf{n}\right) w_{i}\ast w_{j}  \notag
\end{eqnarray}%
where the last line introduces a compact notation for the convolution in
which the spatial arguments are suppressed and where $\mathbf{n}=\lim_{\rho
\left( \mathbf{r}\right) \rightarrow \rho }\mathbf{n}\left( \mathbf{r}%
\right) $.

\subsection{Useful formulae}

For later use, I note that the convolutions are given by%
\begin{align}
w_{\eta }\ast w_{\eta }& =2\pi \Theta \left( \sigma -r_{12}\right) \sigma
^{3}\frac{1}{24}\left( x-1\right) ^{2}\left( x+2\right) \\
w_{s}\ast w_{\eta }& =2\pi \Theta \left( \sigma -r_{12}\right) \sigma ^{2}%
\frac{1}{4}\left( 1-x\right)  \notag \\
w_{s}\ast w_{s}& =2\pi \Theta \left( \sigma -r_{12}\right) \frac{\sigma }{4x}
\notag \\
w_{v_{i}}w_{v_{i}}& =2\pi \Theta \left( \sigma -r_{12}\right) \sigma \left( 
\frac{1}{4x}-\frac{1}{2}x\right)  \notag
\end{align}%
where $x \equiv r_{12}/\sigma$. The weighted densities in the bulk limit are%
\begin{align}
\eta \left( \mathbf{r}\right) & \rightarrow \frac{4\pi }{3}\left( \frac{%
\sigma }{2}\right) ^{3}\rho =\frac{\pi }{6}\rho \sigma ^{3} \\
s\left( \mathbf{r}\right) & \rightarrow 4\pi \left( \frac{\sigma }{2}\right)
^{2}\rho =\frac{6}{\sigma }\eta  \notag \\
v_{i}\left( \mathbf{r}\right) & \rightarrow 0  \notag
\end{align}

\subsection{Rosenfeld functional}

\bigskip It is straightforward to see that the Rosenfeld functional gives%
\begin{align}
\lim_{\rho \left( \mathbf{r}\right) \rightarrow \overline{\rho }}\frac{%
\delta ^{2}\beta F_{ex}^{(R)}}{\delta \rho \left( \mathbf{r}_{1}\right)
\delta \rho \left( \mathbf{r}_{2}\right) }& =s\Phi _{1}^{\prime \prime
}\left( \eta \right) w_{\eta }\ast w_{\eta }+2\Phi _{1}^{\prime }\left( \eta
\right) w_{s}\ast w_{\eta } \\
& +\Phi _{2}^{\prime \prime }\left( \eta \right) \left( s^{2}-v^{2}\right)
w_{\eta }\ast w_{\eta }+2\Phi _{2}\left( \eta \right) \left( w_{s}\ast
w_{s}-w_{v_{i}}\ast w_{v_{i}}\right) +2\Phi _{2}^{\prime }\left( \eta
\right) w_{\eta }\ast \left( 2sw_{s}-2v_{i}w_{v_{i}}\right)  \notag \\
& +\Phi _{3}^{\prime \prime }\left( \eta \right) s\left( s^{2}-3v^{2}\right)
w_{\eta }\ast w_{\eta }+2\Phi _{3}^{\prime }\left( \eta \right) \left(
s^{2}-3v^{2}\right) w_{\eta }\ast w_{s}+2\Phi _{3}^{\prime }\left( \eta
\right) sw_{\eta }\ast 2\left( sw_{s}-3v_{i}w_{v_{i}}\right)  \notag \\
& +2\Phi _{3}\left( \eta \right) w_{s}\ast 2\left(
sw_{s}-3v_{i}w_{v_{i}}\right) +\Phi _{3}\left( \eta \right) s2\left(
w_{s}\ast w_{s}-3w_{v_{i}}\ast w_{v_{i}}\right)  \notag
\end{align}%
and since $v_{i}=0$ in the bulk this becomes%
\begin{align}
\lim_{\rho \left( \mathbf{r}\right) \rightarrow \overline{\rho }}\frac{%
\delta ^{2}\beta F_{ex}^{(R)}}{\delta \rho \left( \mathbf{r}_{1}\right)
\delta \rho \left( \mathbf{r}_{2}\right) }& =s\Phi _{1}^{\prime \prime
}\left( \eta \right) w_{\eta }\ast w_{\eta }+2\Phi _{1}^{\prime }\left( \eta
\right) w_{s}\ast w_{\eta } \\
& +\Phi _{2}^{\prime \prime }\left( \eta \right) s^{2}w_{\eta }\ast w_{\eta
}+2\Phi _{2}\left( \eta \right) \left( w_{s}\ast w_{s}-w_{v_{i}}\ast
w_{v_{i}}\right) +4\Phi _{2}^{\prime }\left( \eta \right) sw_{\eta }\ast
w_{s}  \notag \\
& +\Phi _{3}^{\prime \prime }\left( \eta \right) s\left( s^{2}-3v^{2}\right)
w_{\eta }\ast w_{\eta }+2\Phi _{3}^{\prime }\left( \eta \right) \left(
s^{2}-3v^{2}\right) w_{\eta }\ast w_{s}+4\Phi _{3}^{\prime }\left( \eta
\right) s^{2}w_{\eta }\ast w_{s}  \notag \\
& +6\Phi _{3}\left( \eta \right) s\left( w_{s}\ast w_{s}-w_{v_{i}}\ast
w_{v_{i}}\right)  \notag
\end{align}%
Making use of 
\begin{align}
\Phi _{1}\left( \eta \right) & =-\frac{1}{\pi \sigma ^{2}}\ln \left( 1-\eta
\right) \Longrightarrow \Phi _{1}^{\prime }\left( \eta \right) =\frac{1}{\pi
\sigma ^{2}}\frac{1}{\left( 1-\eta \right) }\Longrightarrow \Phi
_{1}^{\prime \prime }\left( \eta \right) =\frac{1}{\pi \sigma ^{2}}\frac{1}{%
\left( 1-\eta \right) ^{2}} \\
\Phi _{2}\left( \eta \right) & =\frac{1}{2\pi \sigma }\frac{1}{\left( 1-\eta
\right) }\Longrightarrow \Phi _{2}^{\prime }\left( \eta \right) =\frac{1}{%
2\pi \sigma }\frac{1}{\left( 1-\eta \right) ^{2}}\Longrightarrow \Phi
_{2}^{\prime \prime }\left( \eta \right) =\frac{2}{2\pi \sigma }\frac{1}{%
\left( 1-\eta \right) ^{3}}  \notag \\
\Phi _{3}\left( \eta \right) & =\frac{1}{24\pi }\frac{1}{\left( 1-\eta
\right) ^{2}}\Longrightarrow \Phi _{3}^{\prime }\left( \eta \right) =\frac{2%
}{24\pi }\frac{1}{\left( 1-\eta \right) ^{3}}\Longrightarrow \Phi
_{3}^{\prime \prime }\left( \eta \right) =\frac{6}{24\pi }\frac{1}{\left(
1-\eta \right) ^{4}}  \notag
\end{align}
gives%
\begin{eqnarray}
\lim_{\rho \left( \mathbf{r}\right) \rightarrow \overline{\rho }}\frac{%
\delta ^{2}\beta F_{ex}^{(R)}}{\delta \rho \left( \mathbf{r}_{1}\right)
\delta \rho \left( \mathbf{r}_{2}\right) } &=&\frac{1}{2}\frac{\eta }{\left(
\eta -1\right) ^{2}}\left( x-1\right) ^{2}\left( x+2\right) -\frac{1}{\eta -1%
}\left( 1-x\right) \\
&&+-3\frac{\eta ^{2}}{\left( \eta -1\right) ^{3}}\left( x-1\right)
^{2}\left( x+2\right) -\frac{x}{\eta -1}+6\frac{\eta }{\left( \eta -1\right)
^{2}}\left( 1-x\right)  \notag \\
&&+\frac{9}{2}\frac{\eta ^{3}}{\left( \eta -1\right) ^{4}}\left( x-1\right)
^{2}\left( x+2\right) +9\frac{\eta ^{2}}{\left( \eta -1\right) ^{3}}\left(
x-1\right)  \notag \\
&&+\frac{3}{2}x\frac{\eta }{\left( \eta -1\right) ^{2}}  \notag
\end{eqnarray}%
or%
\begin{equation}
\lim_{\rho \left( \mathbf{r}\right) \rightarrow \overline{\rho }}\frac{%
\delta ^{2}\beta F_{ex}^{(R)}}{\delta \rho \left( \mathbf{r}_{1}\right)
\delta \rho \left( \mathbf{r}_{2}\right) }=\allowbreak \frac{1}{2}\frac{\eta
\left( 2\eta +1\right) ^{2}}{\left( 1-\eta \right) ^{4}}x^{3}-\frac{3}{2}%
\frac{\eta \left( \eta +2\right) }{\left( 1-\eta \right) ^{4}}%
^{2}\allowbreak x+\frac{\left( 2\eta +1\right) ^{2}}{\left( 1-\eta \right)
^{4}}
\end{equation}%
$\allowbreak $which is the usual PY expression.

\subsection{The new term}

Next, we need the additional terms introduced by Santos. First note that%
\begin{align}
\frac{\delta \beta \Phi_{S}}{\delta \rho \left( \mathbf{r}_{1}\right) }&
=2\left( sw_{s}-v_{i}w_{v_{i}}\right) \Phi _{2}\left( \eta \right) \Delta \\
& +\left( s^{2}-v^{2}\right) \Phi _{2}^{\prime }\left( \eta \right) w_{\eta
}\Delta  \notag \\
& +\left( s^{2}-v^{2}\right) \Phi _{2}\left( \eta \right) \left( \Delta
_{s}w_{s}+\Delta _{v_{i}}w_{v_{i}}+\Delta _{\eta }w_{\eta }\right)  \notag
\end{align}%
So, neglecting some non-contributing terms proportional to $v_{i}$, 
\begin{align}
\lim_{\rho \left( \mathbf{r}\right) \rightarrow \overline{\rho }}\frac{%
\delta ^{2}\beta \Delta F_{ex}^{(S)}}{\delta \rho \left( \mathbf{r}_{1}\right)
\delta \rho \left( \mathbf{r}_{2}\right) }& =2\left( w_{s}\ast
w_{s}-w_{v_{i}}\ast w_{v_{i}}\right) \Phi _{2}\left( \eta \right) \Delta
+4s\left( w_{s}\ast w_{\eta }\right) \Phi _{2}^{\prime }\left( \eta \right)
\Delta \\
& +4s\Phi _{2}\left( \eta \right) \ast \left( \Delta _{s}w_{s}\ast
w_{s}+\Delta _{\eta }w_{s}\ast w_{\eta }\right) +s^{2}\Phi _{2}^{\prime
\prime }\left( \eta \right) w_{\eta }\ast w_{\eta }\Delta  \notag \\
& +2s^{2}\Phi _{2}^{\prime }\left( \eta \right) \left( \Delta _{s}w_{\eta
}\ast w_{s}+\Delta _{\eta }w_{\eta }\ast w_{\eta }\right)  \notag \\
& +s^{2}\Phi _{2}\left( \eta \right) \left( \Delta _{ss}w_{s}\ast
w_{s}+2\Delta _{s\eta }w_{s}\ast w_{\eta }+2\Delta
_{v^{2}}w_{v_{i}}w_{v_{i}}+\Delta _{\eta \eta }w_{\eta }\ast w_{\eta }\right)
\notag
\end{align}%
Using%
\begin{align}
\Delta & =\Delta \left( 2\frac{s\left( s^{2}-3v^{2}\right) }{\left(
s^{2}-v^{2}\right) }\frac{\Phi _{3}\left( \eta \right) }{\Phi _{2}\left(
\eta \right) }\right) \\
\Delta _{s}& =\left( 2\frac{\left( s^{4}+3v^{4}\right) }{\left(
s^{2}-v^{2}\right) ^{2}}\frac{\Phi _{3}\left( \eta \right) }{\Phi _{2}\left(
\eta \right) }\right) \Delta ^{\prime }\left( 2\frac{s\left(
s^{2}-3v^{2}\right) }{\left( s^{2}-v^{2}\right) }\frac{\Phi _{3}\left( \eta
\right) }{\Phi _{2}\left( \eta \right) }\right)  \notag \\
\Delta _{\eta }& =\left( 2\frac{s\left( s^{2}-3v^{2}\right) }{\left(
s^{2}-v^{2}\right) }\left( \frac{\Phi _{3}\left( \eta \right) }{\Phi
_{2}\left( \eta \right) }\right) ^{\prime }\right) \Delta ^{\prime }\left( 2%
\frac{s\left( s^{2}-3v^{2}\right) }{\left( s^{2}-v^{2}\right) }\frac{\Phi
_{3}\left( \eta \right) }{\Phi _{2}\left( \eta \right) }\right)  \notag \\
\Delta _{v^{2}}& =\left( -4\frac{s^{3}}{\left( s^{2}-v^{2}\right) ^{2}}\frac{%
\Phi _{3}\left( \eta \right) }{\Phi _{2}\left( \eta \right) }\right) \Delta
^{\prime }\left( 2\frac{s\left( s^{2}-3v^{2}\right) }{\left(
s^{2}-v^{2}\right) }\frac{\Phi _{3}\left( \eta \right) }{\Phi _{2}\left(
\eta \right) }\right)  \notag \\
\Delta _{ss}& =\left( 2\frac{\left( s^{4}+3v^{4}\right) }{\left(
s^{2}-v^{2}\right) ^{2}}\frac{\Phi _{3}\left( \eta \right) }{\Phi _{2}\left(
\eta \right) }\right) ^{2}\Delta ^{\prime \prime }\left( 2\frac{s\left(
s^{2}-3v^{2}\right) }{\left( s^{2}-v^{2}\right) }\frac{\Phi _{3}\left( \eta
\right) }{\Phi _{2}\left( \eta \right) }\right)  \notag \\
& +\left( -8s\frac{v^{2}\left( s^{2}+3v^{2}\right) }{\left(
s^{2}-v^{2}\right) ^{3}}\frac{\Phi _{3}\left( \eta \right) }{\Phi _{2}\left(
\eta \right) }\right) \Delta ^{\prime }\left( 2\frac{s\left(
s^{2}-3v^{2}\right) }{\left( s^{2}-v^{2}\right) }\frac{\Phi _{3}\left( \eta
\right) }{\Phi _{2}\left( \eta \right) }\right)  \notag
\end{align}%
and the bulk limits%
\begin{align}
\Delta & \rightarrow \Delta \left( 2s\frac{\Phi _{3}\left( \eta \right) }{%
\Phi _{2}\left( \eta \right) }\right) \\
\Delta _{s}& \rightarrow \left( 2\frac{\Phi _{3}\left( \eta \right) }{\Phi
_{2}\left( \eta \right) }\right) \Delta ^{\prime }\left( 2s\frac{\Phi
_{3}\left( \eta \right) }{\Phi _{2}\left( \eta \right) }\right)  \notag \\
\Delta _{\eta }& \rightarrow 2s\left( \frac{\Phi _{3}\left( \eta \right) }{%
\Phi _{2}\left( \eta \right) }\right) ^{\prime }\Delta ^{\prime }\left( 2s%
\frac{\Phi _{3}\left( \eta \right) }{\Phi _{2}\left( \eta \right) }\right) 
\notag \\
\Delta _{v^{2}}& \rightarrow \left( -\frac{4}{s}\frac{\Phi _{3}\left( \eta
\right) }{\Phi _{2}\left( \eta \right) }\right) \Delta ^{\prime }\left( 2s%
\frac{\Phi _{3}\left( \eta \right) }{\Phi _{2}\left( \eta \right) }\right) 
\notag \\
\Delta _{ss}& \rightarrow \left( 2\frac{\Phi _{3}\left( \eta \right) }{\Phi
_{2}\left( \eta \right) }\right) ^{2}\Delta ^{\prime \prime }\left( 2s\frac{%
\Phi _{3}\left( \eta \right) }{\Phi _{2}\left( \eta \right) }\right)  \notag
\end{align}%
gives%
\begin{align}
\lim_{\rho \left( \mathbf{r}\right) \rightarrow \overline{\rho }}\frac{%
\delta ^{2}\beta \Delta F_{ex}^{(S)}}{\delta \rho \left( \mathbf{r}_{1}\right)
\delta \rho \left( \mathbf{r}_{2}\right) }& =2\left( w_{s}\ast
w_{s}-w_{v_{i}}\ast w_{v_{i}}\right) \Phi _{2}\left( \eta \right) \Delta
+4s\left( w_{s}\ast w_{\eta }\right) \Phi _{2}^{\prime }\left( \eta \right)
\Delta \\
& +4s\Phi _{2}\left( \eta \right) \left( \Delta _{s}w_{s}\ast w_{s}+\Delta
_{\eta }w_{s}\ast w_{\eta }\right) +s^{2}\Phi _{2}^{\prime \prime }\left(
\eta \right) w_{\eta }\ast w_{\eta }\Delta  \notag \\
& +2s^{2}\Phi _{2}^{\prime }\left( \eta \right) \left( \Delta _{s}w_{\eta
}\ast w_{s}+\Delta _{\eta }w_{\eta }\ast w_{\eta }\right)  \notag \\
& +s^{2}\Phi _{2}\left( \eta \right) \left( \left( 2\frac{\Phi _{3}\left(
\eta \right) }{\Phi _{2}\left( \eta \right) }\right) ^{2}\Delta ^{\prime
\prime }w_{s}\ast w_{s}+2\Delta _{s\eta }w_{s}\ast w_{\eta }+\Delta _{\eta
\eta }w_{\eta }\ast w_{\eta }\right) +s^{2}\Phi _{2}\left( \eta \right)
\Delta ^{\prime }\left( -\frac{8}{s}\frac{\Phi _{3}\left( \eta \right) }{%
\Phi _{2}\left( \eta \right) }w_{v_{i}}w_{v_{i}}\right)  \notag
\end{align}%
or, rearranging a little,%
\begin{align}
\lim_{\rho \left( \mathbf{r}\right) \rightarrow \overline{\rho }}\frac{%
\delta ^{2}\beta \Delta F_{ex}^{(S)}}{\delta \rho \left( \mathbf{r}_{1}\right)
\delta \rho \left( \mathbf{r}_{2}\right) }& =\left( 2\left( w_{s}\ast
w_{s}-w_{v_{i}}\ast w_{v_{i}}\right) \Phi _{2}\left( \eta \right) +4s\left(
w_{s}\ast w_{\eta }\right) \Phi _{2}^{\prime }\left( \eta \right) +w_{\eta
}\ast w_{\eta }s^{2}\Phi _{2}^{\prime \prime }\left( \eta \right) \right)
\Delta \\
& +4s\Phi _{2}\left( \eta \right) \left( 2\frac{\Phi _{3}\left( \eta \right) 
}{\Phi _{2}\left( \eta \right) }\left( w_{s}\ast
w_{s}-w_{v_{i}}w_{v_{i}}\right) +3s\left( \frac{\Phi _{3}\left( \eta \right) 
}{\Phi _{2}\left( \eta \right) }\right) ^{\prime }w_{s}\ast w_{\eta
}+s^{2}\left( \frac{\Phi _{3}\left( \eta \right) }{\Phi _{2}\left( \eta
\right) }\right) ^{\prime }w_{\eta }\ast w_{\eta }\right) \Delta ^{\prime } 
\notag \\
& +s^{2}\Phi _{2}\left( \eta \right) \left( \left( 2\frac{\Phi _{3}\left(
\eta \right) }{\Phi _{2}\left( \eta \right) }\right) ^{2}w_{s}\ast
w_{s}\Delta ^{\prime \prime }+2\Delta _{s\eta }w_{s}\ast w_{\eta }+\Delta
_{\eta \eta }w_{\eta }\ast w_{\eta }\right)  \notag
\end{align}%
Next, using%
\begin{align}
\Delta _{\eta s}& =\left( 2\frac{s\left( s^{2}-3v^{2}\right) }{\left(
s^{2}-v^{2}\right) }\left( \frac{\Phi _{3}\left( \eta \right) }{\Phi
_{2}\left( \eta \right) }\right) ^{\prime }\right) \left( 2\frac{\left(
s^{4}+3v^{4}\right) }{\left( s^{2}-v^{2}\right) ^{2}}\frac{\Phi _{3}\left(
\eta \right) }{\Phi _{2}\left( \eta \right) }\right) \Delta ^{\prime \prime
}\left( 2\frac{s\left( s^{2}-3v^{2}\right) }{\left( s^{2}-v^{2}\right) }%
\frac{\Phi _{3}\left( \eta \right) }{\Phi _{2}\left( \eta \right) }\right) \\
& +\left( 2\frac{\left( s^{4}+3v^{4}\right) }{\left( s^{2}-v^{2}\right) ^{2}}%
\left( \frac{\Phi _{3}\left( \eta \right) }{\Phi _{2}\left( \eta \right) }%
\right) ^{\prime }\right) \Delta ^{\prime }\left( 2\frac{s\left(
s^{2}-3v^{2}\right) }{\left( s^{2}-v^{2}\right) }\frac{\Phi _{3}\left( \eta
\right) }{\Phi _{2}\left( \eta \right) }\right)  \notag \\
& \rightarrow 4s\left( \frac{\Phi _{3}\left( \eta \right) }{\Phi _{2}\left(
\eta \right) }\right) ^{\prime }\left( \frac{\Phi _{3}\left( \eta \right) }{%
\Phi _{2}\left( \eta \right) }\right) \Delta ^{\prime \prime }+2\left( \frac{%
\Phi _{3}\left( \eta \right) }{\Phi _{2}\left( \eta \right) }\right)
^{\prime }\Delta ^{\prime }  \notag \\
\Delta _{\eta \eta }& =\left( 2\frac{s\left( s^{2}-3v^{2}\right) }{\left(
s^{2}-v^{2}\right) }\left( \frac{\Phi _{3}\left( \eta \right) }{\Phi
_{2}\left( \eta \right) }\right) ^{\prime }\right) ^{2}\Delta ^{\prime
\prime }\left( 2\frac{s\left( s^{2}-3v^{2}\right) }{\left(
s^{2}-v^{2}\right) }\frac{\Phi _{3}\left( \eta \right) }{\Phi _{2}\left(
\eta \right) }\right)  \notag \\
& +\left( 2\frac{s\left( s^{2}-3v^{2}\right) }{\left( s^{2}-v^{2}\right) }%
\left( \frac{\Phi _{3}\left( \eta \right) }{\Phi _{2}\left( \eta \right) }%
\right) ^{\prime \prime }\right) \Delta ^{\prime }\left( 2\frac{s\left(
s^{2}-3v^{2}\right) }{\left( s^{2}-v^{2}\right) }\frac{\Phi _{3}\left( \eta
\right) }{\Phi _{2}\left( \eta \right) }\right)  \notag \\
& \rightarrow 4s^{2}\left( \left( \frac{\Phi _{3}\left( \eta \right) }{\Phi
_{2}\left( \eta \right) }\right) ^{\prime }\right) ^{2}\Delta ^{\prime
\prime }+2s\left( \frac{\Phi _{3}\left( \eta \right) }{\Phi _{2}\left( \eta
\right) }\right) ^{\prime \prime }\Delta ^{\prime }  \notag
\end{align}%
results in
\begin{align}
& \lim_{\rho \left( \mathbf{r}\right) \rightarrow \overline{\rho }}\frac{%
\delta ^{2}\beta \Delta F_{ex}^{(S)}}{\delta \rho \left( \mathbf{r}_{1}\right)
\delta \rho \left( \mathbf{r}_{2}\right) } \\
& =\left( 2\left( w_{s}\ast
w_{s}-w_{v_{i}}\ast w_{v_{i}}\right) \Phi _{2}\left( \eta \right) +4s\left(
w_{s}\ast w_{\eta }\right) \Phi _{2}^{\prime }\left( \eta \right) +w_{\eta
}\ast w_{\eta }s^{2}\Phi _{2}^{\prime \prime }\left( \eta \right) \right)
\Delta \notag \\
& +4s\Phi _{2}\left( \eta \right) \left( 2\frac{\Phi _{3}\left( \eta \right) 
}{\Phi _{2}\left( \eta \right) }\left( w_{s}\ast
w_{s}-w_{v_{i}}w_{v_{i}}\right) +4s\left( \frac{\Phi _{3}\left( \eta \right) 
}{\Phi _{2}\left( \eta \right) }\right) ^{\prime }w_{s}\ast w_{\eta
}+s^{2}\left( \left( \frac{\Phi _{3}\left( \eta \right) }{\Phi _{2}\left(
\eta \right) }\right) ^{\prime }+\frac{1}{2}\left( \frac{\Phi _{3}\left(
\eta \right) }{\Phi _{2}\left( \eta \right) }\right) ^{\prime \prime
}\right) w_{\eta }\ast w_{\eta }\right) \Delta ^{\prime }  \notag \\
& +s^{2}\Phi _{2}\left( \eta \right) \left( \left( 2\frac{\Phi _{3}\left(
\eta \right) }{\Phi _{2}\left( \eta \right) }\right) ^{2}w_{s}\ast
w_{s}+8s\left( \frac{\Phi _{3}\left( \eta \right) }{\Phi _{2}\left( \eta
\right) }\right) ^{\prime }\left( \frac{\Phi _{3}\left( \eta \right) }{\Phi
_{2}\left( \eta \right) }\right) w_{s}\ast w_{\eta }+4s^{2}\left( \left( 
\frac{\Phi _{3}\left( \eta \right) }{\Phi _{2}\left( \eta \right) }\right)
^{\prime }\right) ^{2}w_{\eta }\ast w_{\eta }\right) \Delta ^{\prime \prime }
\notag
\end{align}%
Inserting the convolutions and the bulk value of $s$ gives%
\begin{align}
& \lim_{\rho \left( \mathbf{r}\right) \rightarrow \rho }\frac{\delta
^{2}\beta \Delta F_{ex}^{(S)}}{\delta \rho \left( \mathbf{r}_{1}\right) \delta
\rho \left( \mathbf{r}_{2}\right) }  =  \Theta(\sigma-r_{12}) K(r_{12}/\sigma) \\
&K(x) =2\pi \sigma \left( x\Phi _{2}\left( \eta \right) +6\eta \left( 1-x\right)
\Phi _{2}^{\prime }\left( \eta \right) +\frac{3}{2}\eta ^{2}\left(
x-1\right) ^{2}\left( x+2\right) \Phi _{2}^{\prime \prime }\left( \eta
\right) \right) \Delta  \notag \\
& +48\pi \eta \Phi _{2}\left( \eta \right) \left( \frac{\Phi _{3}\left( \eta
\right) }{\Phi _{2}\left( \eta \right) }x+6\eta \left( \frac{\Phi _{3}\left(
\eta \right) }{\Phi _{2}\left( \eta \right) }\right) ^{\prime }\left(
1-x\right) +\frac{3}{2}\eta ^{2}\left( \left( \frac{\Phi _{3}\left( \eta
\right) }{\Phi _{2}\left( \eta \right) }\right) ^{\prime }+\frac{1}{2}\left( 
\frac{\Phi _{3}\left( \eta \right) }{\Phi _{2}\left( \eta \right) }\right)
^{\prime \prime }\right) \left( x-1\right) ^{2}\left( x+2\right) \right)
\Delta ^{\prime }  \notag \\
& +72\frac{\pi }{\sigma }\eta ^{2}\Phi _{2}\left( \eta \right) \left( \left( 
\frac{\Phi _{3}\left( \eta \right) }{\Phi _{2}\left( \eta \right) }\right)
^{2}\frac{1}{x}+12\eta \left( \frac{\Phi _{3}\left( \eta \right) }{\Phi
_{2}\left( \eta \right) }\right) ^{\prime }\left( \frac{\Phi _{3}\left( \eta
\right) }{\Phi _{2}\left( \eta \right) }\right) \left( 1-x\right) +6\eta
^{2}\left( \left( \frac{\Phi _{3}\left( \eta \right) }{\Phi _{2}\left( \eta
\right) }\right) ^{\prime }\right) ^{2}\left( x-1\right) ^{2}\left(
x+2\right) \right) \Delta ^{\prime \prime }  \notag
\end{align}%
Finally, using the explicit forms for $\Phi _{2}\left( \eta \right) $ and $\Phi
_{3}\left( \eta \right) $ and noting that 
\begin{align}
\frac{\Phi _{3}\left( \eta \right) }{\Phi _{2}\left( \eta \right) }& =\frac{1%
}{12}\frac{\sigma }{1-\eta } \\
2s\frac{\Phi _{3}\left( \eta \right) }{\Phi _{2}\left( \eta \right) }& =%
\frac{\eta }{1-\eta }  \notag
\end{align}%
results in the form given in the main text.%

\bibliography{comment_as}

\end{document}